\documentclass[journal]{IEEEtran}
\IEEEoverridecommandlockouts
\usepackage{cite}
\usepackage{amsmath,amssymb,amsfonts}
\usepackage{algorithm}
\usepackage{algorithm,algpseudocode}
\usepackage{graphicx}
\usepackage{multirow}
\usepackage{caption}
\usepackage{subcaption}
\usepackage{textcomp}
\usepackage{xcolor}
\usepackage{amsmath, amsthm}
\def\BibTeX{{\rm B\kern-.05em{\sc i\kern-.025em b}\kern-.08em
    T\kern-.1667em\lower.7ex\hbox{E}\kern-.125emX}}

\begin{document}
\title{Discrete Phase-Shift Optimization of Reconfigurable Intelligent Surfaces using Uniform Reference Directions} 

\author{Pavithra P$^{\dagger}$, Manoj Kumar Panda$^{\dagger,*}$ and Chandramani Singh$^{@}$ 
\thanks{$^*$Corresponding Author}
\thanks{$^{\dagger}$Pavithra P and Manoj Kumar Panda are with the Department of Electronics and Communication Engineering, Amrita School of Engineering, Bengaluru, Amrita Vishwa Vidyapeetham, India. Email: ppavithra.erode@gmail.com, kp\_manoj@blr.amrita.edu}
\thanks{$^{@}$Chandramani Singh is with the Department of Electronic Systems Engineering (DESE), Indian Institute of Science (IISc), Bangalore, Karnataka, India. Email: chandra@iisc.ac.in} 
}

\maketitle

\begin{abstract}
Reconfigurable Intelligent Surfaces (RISs) enable programmable control of the wireless propagation environment through adjustable reflecting elements. A fundamental challenge in RIS-assisted communications is the optimization of discrete phase shifts to maximize the received signal power. While recently developed algorithms can obtain optimal solutions with linear complexity, their computational cost may still be excessive in applications requiring large-scale simulations, real-time adaptation, or repeated optimization over numerous channel realizations. This paper proposes a low-complexity discrete phase-shift optimization algorithm based on multiple reference directions within the feasible optimality sector. By evaluating candidate phase configurations generated from uniformly distributed reference directions, the proposed method systematically improves upon existing approximation approaches that rely on a fixed set of reference directions. The number of reference directions serves as a design parameter that enables a controllable trade-off between computational complexity and optimality. Extensive simulations demonstrate that the proposed algorithm achieves substantial computational speedups while maintaining a small optimality gap relative to the optimal solution. The results further show that increasing the number of reference directions progressively reduces the optimality gap, providing flexible operating points that can be tailored to application-specific complexity and performance requirements. 
\end{abstract}

\begin{IEEEkeywords}
Reconfigurable Intelligent Surface (RIS), Intelligent Reflecting Surface (IRS), Phase Shift Optimization, Unit-Gain Amplitude, Geometric Interpretation, Sector-Based Beamforming, Low-Complexity Algorithms, Received Power, Directional Beamforming.
\end{IEEEkeywords}

\section{Introduction} 

\begin{figure}[ht]
    \centering
    \includegraphics[width=0.48\textwidth]{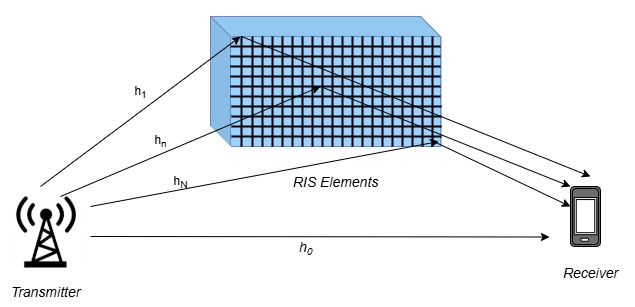}
    \caption{An RIS-aided Communication System. Here, an RIS panel consisting of a rectangular array of $N$ REs connects a transmitter and a receiver through $N$ Non-Line-of-Sight (NLoS) paths having complex channel gains, $h_n$, $1 \leq n \leq N$, where the $n$-th RE can introduce an additional phase shift of $\theta_n$. Optionally, there can be a direct Line-of-Sight (LoS) path having a complex channel gain of $h_0$. } 
     \label{fig:RIS}
\end{figure}

Reconfigurable Intelligent Surface (RIS), also known as Intelligent Reflecting Surface (IRS), is a disruptive technology proposed for beyond 5G wireless communications \cite{chowdhury2020}. Unlike conventional wireless communication systems, an RIS-aided system can effectively modify the propagation environment itself, thereby providing a new degree of freedom to optimize the performance \cite{renzo2019}. By adjusting the phase shifts of a large array of nearly passive \textit{reflecting elements} (REs), an RIS can enhance signal strengths at targeted locations without the need for dedicated power amplifiers. See Fig. \ref{fig:RIS}. 

RISs provide a promising solution to improve the performance of next-generation wireless communications. However, there are some concerns that need to be addressed. First, a significant challenge is to jointly optimize the phase-shifts of a large number of REs in order to achieve a certain well-defined objective, e.g., maximization of channel capacity or received signal to noise ratio (SNR), minimization of transmit power, etc. Second, due to hardware constraints, each RE can introduce only a set of \textit{discrete} phase-shift values, which results in a large-scale discrete and non-convex optimization problem \cite{bjornson2022,wu2021}. An excellent survey of different optimization formulations, covering a wide variety of objective functions and with discrete as well as continuous phase-shifts, and their corresponding solution approaches can be found in \cite{zhou2023}. 

In this paper, we focus on the problem of maximization of received SNR using discrete phase shifts. Recent studies have investigated the cases of \textit{nonuniform} or unequally-spaced discrete phase-shifts \cite{hashemi2024a,pekan2024b}, and where the REs can introduce \textit{phase-shift-dependent amplitude gains} \cite{hashemi2026}. However, as in \cite{zhang2022,ren2023,zhao2024,pekan2024a}, we focus on the case where the REs introduce \textit{uniform} or equally-spaced discrete phase-shifts and \textit{unit-amplitude gains}. Although the considered setting is simpler, the resulting optimization problem is still computationally challenging because the cardinality of the feasible set of phase-shift vectors increases \textit{exponentially} with the number of REs as $K^N$, where $N$ denotes the numbers of REs in the RIS panel and $K$ denotes the number of phase shifts that any RE can introduce. 

For the considered problem, the authors in \cite{ren2023,zhao2024,pekan2024a} have proposed \textit{linear time} algorithms that obtain the optimal phase-shifts with an asymptotic complexity of $O(N)$. Our goal, however, is to improve the computational speed for practical values of $N$ beyond the `linear time' algorithms in \cite{ren2023,zhao2024,pekan2024a} at the expense of a negligible `optimality gap', i.e., a negligible percentage of reduction in the received SNR relative to the optimal (maximum) value. 

The contributions of the paper are as follows: 
\begin{enumerate}
    \item[\textbf{C1}.] We develop a sub-optimal algorithm, the \textbf{Uniform Reference Directions (URD)} algorithm, parameterized by the number of reference directions $D$. Existing approximation algorithms including CPP, APX and ARB arise as special cases corresponding to particular values of DDD. The sub-optimal algorithm ARB in \cite{pavithra2025} is a special case of the URD algorithm with $D\,{=}\,1$. Also, the sub-optimal algorithm Closest Point Projection (CPP) \cite{zhou2023} and the Approximation (APX) algorithm in \cite{zhang2022} are equivalent to URD with ${D\,{=}\,1}$ and ${D\,{=}\,3}$, respectively. The corresponding `speed-ups' are, however, found to be slightly higher for the URD algorithm due to its simplicity of implementation. 
    \item[\textbf{C2}.] We conduct extensive simulations using MATLAB\textsuperscript{\textregistered} and study how the `optimality gap' and `speed-up' vary with the number of REs, $N$, the parameter $D$, and for different number of phase-shifts, $K$. We show that: 
    \begin{itemize} 
    \item As $N$ increases, the \textit{worst-case} `optimality gap' decreases while the \textit{average} `speed-up' increases, making the URD algorithm particularly effective for challenging cases.\footnote{Clearly, the computations required for cases with small $N$ can be easily handled even using exhaustive search, and so the real benefit of the URD algorithm can only be measured in cases with large $N$, where even the `linear time' algorithms may be considered to be slow for certain applications.} 
    \item As the parameter $D$ increases, the `optimality gap' decreases, but the `speed-up' also decreases. Thus, the parameter $D$ can be tuned to strike a balance between `optimality gap' and `speed-up' to match a target application's requirements.\footnote{An application of the URD algorithm could be to generate training dataset for a Deep Learning (DL) model which would be used for DL-based phase-shift optimization. Here, the parameter $D$ can be tuned to trade off between the time taken for training data generation and `optimality gap'.}  
    \item Depending on the values of $K$ and $N$, and $D$ considered, the proposed URD algorithm could provide solutions up to 3-30 times faster, sacrificing only 0.01\%-10\% of optimal SNR in the worst case.     
\end{itemize}
    \item[\textbf{C3}.] We provide a convergence argument showing that the received power achieved by URD approaches the optimal received power as $D$ goes to infinity, causing the corresponding optimality gap to vanish. 
\end{enumerate}

%in the exhaustive search approach is impractical. Note, however, that our `speed-up' results are relative to the `linear time' algorithms reported in \cite{ren2023,zhao2024,pekan2024a} and not w.r.t. exhaustive search. 

%\footnote{Due to space constraints, the applicability of our proposed technique to the more general case of nonuniform phase-shifts and phase-shift-dependent amplitude gains will be demonstrated in a companion paper.} 

The remainder of the paper is organized as follows.

\section{System model} 

%\begin{figure}[ht]
%    \centering
%    \includegraphics[width=0.48\textwidth]{RIS_2.jpg}
%    \caption{An RIS-aided Communication System.}
%     \label{fig:RIS}
%\end{figure}

We consider a wireless communication system, aided by an RIS panel, as shown in Fig.~\ref{fig:RIS}. Both the transmitter and the receiver are equipped with single antennas. Each of the $N$ REs is capable of introducing a phase shift to the incoming signal. The discrete set of possible phase shifts, denoted by $\Phi_K$, is given by 
\begin{equation} 
\label{eq:PhiK}
\Phi_K\,{=}\,\{\omega, 2\omega, \dots, K\omega\}, \quad \text{where} \quad \omega\,{=}\,\frac{2\pi}{K}.  
\end{equation}

Let \(h_{n}\) denote the complex-valued channel gain of the path from the transmitter to the receiver via the $n$-th RE on the RIS panel. Let \(h_{0}\) denote the aggregate channel gain of all remaining paths that do not go via the RIS, including the direct Line of Sight (LoS) path, if any. The gain $h_n$ can be rewritten in the polar form as 
\begin{equation}
\label{eq:hnpolarform} 
    h_n\,{=}\,\beta_n e^{j\alpha_n}, \quad n\,{=}\,0, 1, \dots, N, 
\end{equation}
where \( \beta_n \) denotes the amplitude and \( \alpha_n \) represents the phase of the channel gain of the path via the $n$-th RE. 

Let \({\theta_n\,{\in}\,\Phi_K}\), $1 \leq n \leq N$, denote the phase shift introduced by the $n$-th RE. The received signal, \( {Y\,{\in}\,\mathbb{C}} \), corresponding to the transmitted signal, \( {X\,{\in}\,\mathbb{C}} \), is given by 
\begin{equation} 
\label{eq:RxSignal}
Y\,{=}\,\left(h_{0} + \sum_{n\,{=}\,1}^{N} h_{n} e^{j\theta_n} \right) X + Z, 
\end{equation}
where \( Z \) represents the additive noise present in the system. The aim is to find an optimal value of the phase-shift vector \(\boldsymbol{\theta}\,{=}\,(\theta_1, \dots, \theta_N)\) that yields the maximum received signal power, $E[Y^2]$, for given values of transmit signal power, $E[X^2]$, noise power, $E[Z^2]$, and channel gains, $h_n$, $\forall n$. To that end, we formulate and solve the following optimization problem. \\ 

\noindent \textbf{Problem P:} 
\begin{equation} 
\label{eq:optimizationProblem}
    \begin{array}{ll}
        \text{maximize}  & f(\boldsymbol{\theta}) \\
        \text{subject to} & \theta_n\,{\in}\,\Phi_K, \quad \text{for} \quad n\,{=}\,1, \dots, N, 
    \end{array}
\end{equation}
where the objective function \(f(\boldsymbol{\theta})\) is given by 
\begin{equation}
\label{eq:objFunction}
f(\boldsymbol{\theta})\,{=}\,\left| h_0 + \sum_{n\,{=}\,1}^{N} h_n e^{j\theta_n} \right|^2.
\end{equation}
It is clear from \eqref{eq:optimizationProblem} that the feasible solution space 
\[\underbrace{\Phi_K^N\,{=}\,\Phi_K \times \dots \times \Phi_K}_{N \, \, times}\] 
grows exponentially with \(N\), making exhaustive search impractical.

\section{Methodology} 
\label{sec:methodology} 
%For the considered problem of maximization of received SNR using REs with discrete uniform phase shifts and unit amplitude gains, 

\subsection{From Exponential to Linear Complexity}
In this subsection, we quickly recap the main concepts that have resulted in the reduction of the computational complexity in \cite{zhang2022,ren2023,zhao2024,pekan2024a,pavithra2025} from exponential to linear in $N$. For Problem P, the authors in \cite{zhang2022,ren2023,zhao2024,pekan2024a,pavithra2025} leverage the geometric insight that the maximization of the received SNR is equivalent to maximizing the magnitude of the sum of $N$ two-dimensional vectors, $h_n e^{j\theta_n}$, $1\,{\leq}\,n\,{\leq}\,N$ (see \eqref{eq:objFunction}), and this maximization is achieved by finding optimal rotations, $\theta_n^*$, $1\,{\leq}\,n\,{\leq}\,N$, so as to make the rotated channel gain vectors as close as possible to each other, and thereby maximize the vector sum.\footnote{The channel gain pertaining to each path from the source to the destination through an RE is a complex number and can be viewed as a vector on the two-dimensional complex plane. If the $n$-th RE introduces a phase shift of $\theta_n$, then the corresponding channel gain vector $h_n$ rotates anti-clockwise by an angle $\theta_n$ on the two-dimensional complex plane. Clearly, the set of allowable angles of rotation is identical to the set of discrete phase shifts, $\Phi_K$.}

\begin{table*}[t]
\caption{Comparison of the proposed Uniform Reference Direction (URD) algorithm with existing works. Here, $h_0$ denotes the effective (aggregated) channel gain of all the paths that do not go via any of the $N$ REs, if any such paths exist.}
\label{table:I}
\centering
%\begingroup
%\setlength{\tabcolsep}{4pt} 
%\renewcommand{\arraystretch}{1.5} % Default value: 1
%\begin{tabular}{|m{2cm}|m{2.5cm}|m{3cm}|}
\begin{tabular}{|c|c|c|c|c|}
\hline
\multirow{2}{*}{\textbf{Algorithm}} &  \multirow{2}{*}{\textbf{Solution Type}} & \multicolumn{3}{c|}{\textbf{Time Complexity}} \\
\cline{3-5}
& & \textbf{T1} & \textbf{T2} & \textbf{T3} \\
\hline
%\multirow{2}{*}{\cite{zhang2022}} & Optimal only for $K\,{=}\,2$ & \multicolumn{3}{c|}{Overall $\mathcal{O}(N)$ for $K\,{=}\,2$} \\
%\cline{2-5}
\cite{zhang2022} & \textbf{Algo. APX:} Sub-Optimal, $K\,{>}\,2$ & $\mathcal{O}(1)$ & $\mathcal{O}(1)$ & $\mathcal{O}(3N)$ \\
\hline
\multirow{2}{*}{\cite{ren2023}} & Optimal for $h_0\,{\ne}\,0$ & $\mathcal{O}(2N)$ & \multirow{2}{*}{Avg. $\mathcal{O}(N \log(N))$; Worst Case $\mathcal{O}(N^2)$} & \multirow{2}{*}{$\mathcal{O}(N)$} \\
\cline{2-3} 
& Optimal for $h_0\,{=}\,0$ & $\mathcal{O}(KN)$ & & \\
\hline 
\cite{zhao2024}, \cite{pekan2024a} & Optimal for any $h_0\,{=}\,0$ & $\mathcal{O}(N)$ & Avg. $\mathcal{O}(N \log(N))$; Worst Case $\mathcal{O}(N^2)$ & $\mathcal{O}(N)$ \\
%\hline
%\cite{pekan2024a} & Optimal for any $h_0$ & $\mathcal{O}(NK)$ & Avg. $\mathcal{O}(N \log(N))$; Worst Case $\mathcal{O}(N^2)$ & $\mathcal{O}(N)$ \\
\hline
\multirow{2}{*}{\cite{pavithra2025}} & Optimal for any $h_0$ as $K\,{\rightarrow \infty}\,$ & \multirow{2}{*}{$\mathcal{O}(1)$} & \multirow{2}{*}{$\mathcal{O}(1)$} & \multirow{2}{*}{$\mathcal{O}(N)$} \\
%\cline{2-5}
 & \textbf{Algo. ARB:} Sub-Optimal for $K < \infty$ &  &  & \\
 \hline 
\multirow{2}{*}{This Paper} & Optimal for any $h_0$ as $D\,{\rightarrow \infty}\,$ & \multirow{2}{*}{$\mathcal{O}(D)$} & \multirow{2}{*}{$\mathcal{O}(1)$} & \multirow{2}{*}{$\mathcal{O}(DN)$} \\
%\cline{2-5}
 & \textbf{Algo. URD:} Sub-Optimal for $D < \infty$ & & & \\
 \hline 
\end{tabular}
%\endgroup
\end{table*}

The works \cite{zhao2024,pekan2024a,pavithra2025} combine the geometric insight mentioned above with the rotational symmetry of optimal solutions, assuming the NLoS condition.\footnote{With $K$ uniform (or equally-spaced) phase-shifts and no LoS path, the optimal solutions exhibit rotational symmetry in the sense that if the optimally rotated (or phase-shifted) channel gain vectors pertaining to each RE is further rotated by the same angle, ${\omega\,{=}\,2\pi/K}$, then the resulting rotations are also optimal as the magnitude of the vector sum remains unchanged. See Proposition 1 in \cite{zhao2024}. The NLoS condition, which is quite realistic in high-frequency of operation, e.g., in mmWave or THz bands, implies that ${h_0\,{=}\,0}$. In fact, the absence of a reasonably strong LoS path is precisely what mandates the inclusion of an RIS and create reasonably strong reflected paths.} Let \(\boldsymbol{\theta^*}\,{=}\,(\theta_1^*, \dots, \theta_N^*)\,{\in}\,\Phi_K^N\) represent an optimal vector of phase shifts, i.e., it is a solution for Problem P. Define the unit vector $\mu^*\,{\in}\,\mathbb{C}$ along an optimal sum vector, $\sum_{n\,{=}\,1}^{N} h_n e^{j\theta_n^*}$, to be an \textit{optimal direction}, assuming the NLoS condition. The rotational symmetry of optimal solutions implies that if $\mu^*$ is an optimal direction, then 
\[\mu^*_k\,{=}\,\mu^* e^{jk\omega}\,{=}\,\mu^* e^{j2k\pi/K}, k\,{=}\,1, \dots, {K\,-\,1},\] 
is also an optimal direction. Clearly, there are at least $K$ distinct optimal directions and there is an optimal direction in every ${\omega\,{=}\,2\pi/K}$ sector of the complex plane (see Proposition 1 in \cite{zhao2024}). Lemma 1 in \cite{ren2023} establishes that if, $\forall n$, the complex channel gain vector, $h_n$, is rotated using the discrete phase shifts, $\theta_n\,{\in}\,\Phi_K$, so as to align it as closely as possible with any fixed optimal direction, $\mu^*$, then the resulting rotations are optimal. The main challenge, however, is that an optimal direction is not known \textit{a priori}; it is found only after finding the optimal rotations. 

The main idea in \cite{zhang2022,ren2023,zhao2024,pekan2024a,pavithra2025} is to: (i) carefully construct a set, $\mathcal{M}\,{=}\,\{\mu_1, \dots, \mu_M\}$, of \textit{reference directions} which act as surrogates for an optimal direction, $\mu^*$, (ii) consider one reference direction, $\mu_i\,{\in}\,\mathcal{M}$, at a time to align the channel gain vectors with $\mu_i$ and compute the corresponding phase rotations and the vector sum, and finally (iii) choose the reference direction and the corresponding phase rotations as the solution which maximize the magnitude of the vector sum.

\subsection{From Optimal Algorithms with Linear Time Complexity to Faster Sub-Optimal Algorithms}
The algorithms in \cite{zhang2022,ren2023,zhao2024,pekan2024a,pavithra2025} differ in the ways the following tasks (\textbf{T1}-\textbf{T3}) are performed: 
\begin{enumerate}
\item [\textbf{T1.}] \textbf{Generate:} Construct an appropriate set of reference directions, $\mathcal{M}$, based on the channel gain vectors, and with or without leveraging rotational symmetry. 
\item [\textbf{T2.}] \textbf{Sort:} Sort the reference directions in $\mathcal{M}$ in an appropriate order to be considered sequentially. 
\item [\textbf{T3.}] \textbf{Sweep-Through:}  Rotate an appropriate subset of channel gain vectors for every considered reference direction $\mu_i\,{\in}\,\mathcal{M}$. 
\end{enumerate} 
The computational complexity of the algorithms in \cite{zhang2022,ren2023,zhao2024,pekan2024a,pavithra2025} and whether the resulting solutions are optimal or sub-optimal are determined by the way the tasks (\textbf{T1}-\textbf{T3}) are performed.

\begin{figure}[tbh]
    \centering
    \includegraphics[width=0.4\textwidth]{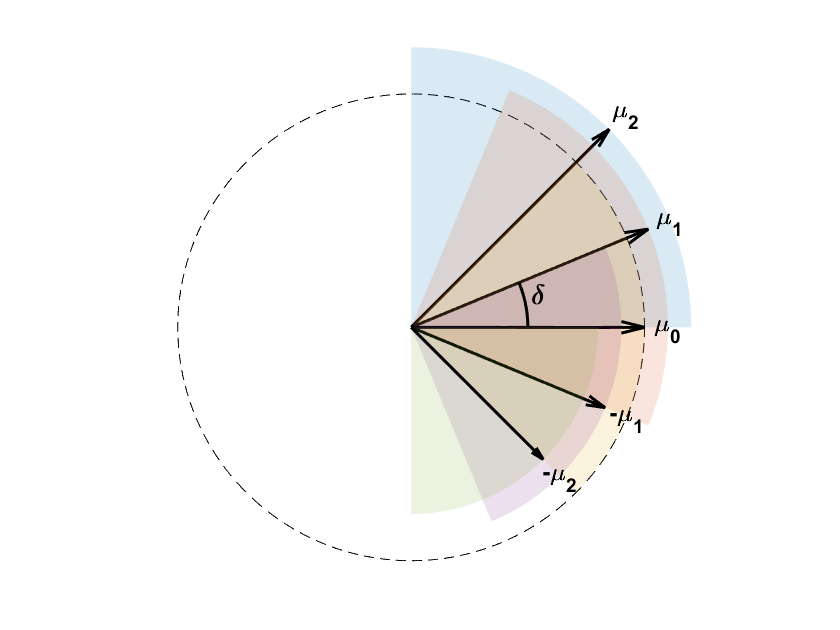}
    \caption{Uniformly Spaced Reference Directions, $\mu_i$, and Their Sectors, $S_i\,{=}\,[\mu_i - \frac{\omega}{2}, \mu_i + \frac{\omega}{2}]$, $-P\,{\leq}\,i\,{\leq}\,P$.}
     \label{fig:sectors}
\end{figure}

A comparison of the computational complexity and optimality characteristics of the algorithms in \cite{zhang2022,ren2023,zhao2024,pekan2024a,pavithra2025} is provided in Table I. It can be observed that, unlike the methods in \cite{zhang2022} and \cite{ren2023}, the algorithms in \cite{zhao2024,pekan2024a,pavithra2025} achieve complexities that are independent of the number of quantized phase shifts, $K$, by exploiting the rotational symmetry of optimal solutions. Among the existing approaches, the sub-optimal algorithm in \cite{pavithra2025} offers the lowest computational complexity and has been reported to incur a negligible average optimality gap (typically below 1\%). However, a small \textit{average} optimality gap does not necessarily imply uniformly good performance across all channel realizations. As demonstrated by our numerical results, the \textit{worst-case} optimality gaps of the sub-optimal algorithms in \cite{zhang2022} and \cite{pavithra2025} can become arbitrarily large, approaching 100\% for certain channel conditions. Furthermore, these algorithms provide only fixed operating points in the complexity-performance trade-off space and do not offer a mechanism to adjust the balance between computational cost and optimality according to application requirements. This observation motivates the development of a more flexible approach that enables a controllable trade-off between computational complexity and solution quality.

\begin{algorithm}[tbh]
\caption{Uniform Reference Direction (URD) Algorithm} 
\label{algo:example}
\begin{algorithmic}[1] % The [1] enables line numbering
    \State \textbf{Input:} Channel vector $\mathbf{h}\,{=}\,[h_0, h_1, \dots, h_N]$, quantization level $K$, and the number of reference directions $D$.
    \State Compute quantization interval $\omega\,{=}\,\frac{2\pi}{K}$ and angular offset $\quad \delta\,{=}\,\frac{\omega}{D-1}$.
    \State Define reference directions: \[\angle \mu_i\,{=}\,\angle h_0 + i\delta, \quad \mbox{for} \quad i \in [-(D-1)/2,(D-1)/2].\] 
    \State Set $f_{\text{max}} = 0$. 
    \For{each reference direction $\mu_i$}
    \State Define sector $S_i\,{=}\,[\mu_i - \omega/2, \mu_i + \omega/2]$. 
    \For{$n\,{=}\,1$ to $N$}
        \State Assign $\theta_n \in \Phi_K$ such that $\angle(h_n e^{j\theta_n}) \in S_i$. 
    \EndFor
    \State Compute $f\,{=}\,\left| h_0 + \sum_{n\,{=}\,1}^N h_n e^{j\theta_n} \right|^2$. 
    \If{$f > f_{\text{max}}$}
        \State Update $f_{\text{max}} \gets f$, $\boldsymbol{\theta}_{\text{opt}} \gets \boldsymbol{\theta}$. 
    \EndIf
    \EndFor
\State \textbf{Output:} Optimal phase shift vector $\boldsymbol{\theta}_{\text{opt}}$. 
\State Compute optimal received power as $|h_0 + \sum h_n e^{j \theta_{\text{opt}}}|^2$. 
\end{algorithmic}
\end{algorithm}

\subsection{The Uniform Reference Directions Algorithm} 

Define a primary reference direction $\mu_0\,{\in}\,\mathbb{C}$ (see Fig. \ref{fig:sectors}). If there exists a path from the transmitter to the receiver that does not go through the RIS, i.e., if $|h_0| \neq 0$, then set $Arg(\mu_0)\,{=}\,Arg(h_0)$. If $|h_0|\,{=}\,0$, then $\mu_0$ can be set to any arbitrary direction. For simplicity, we set it as $Arg(\mu_0)\,{=}\,0$. Define $P$ other reference directions on either side of $\mu_0$ as shown in Fig. \ref{fig:sectors}. The $i$-th reference direction is $\mu_i\,{\in}\,\mathbb{C}$, $-P\,{\leq}\,i\,{\leq}\,P$. The reference direction $\mu_i$, $i\,{\neq}\,0$, makes an angle $i \delta\,{=}\,i \left(\frac{\omega}{2P}\right)$ with the primary reference direction $\mu_0$. The total number of reference directions, $D$, is, therefore, given by $D\,{=}\,2P+1$. Notice that the $D$ reference directions, $\mu_i$, $-P\,{\leq}\,i\,{\leq}\,P$, are \textit{uniformly} spaced, and together, cover a sector of angle $\omega$ within which an optimal direction must lie. Furthermore, they can be determined without the knowledge of channel gains and are \textit{auto-sorted}. Therefore, the complexity of task \textbf{T2} now reduces to $O(1)$, and even though the task \textbf{T1} now involves an asymptotic complexity of $O(D)$, \textit{it can be computed once and then used for all channel realizations}. 

The proposed Uniform Reference Directions (URD) algorithm is shown as Algorithm 1. The main idea is to construct a tunable number, $D$, of uniformly spaced reference directions, consider one reference direction at a time, rotate all channel gain vectors to align as closely as possible with the chosen reference direction, and finally choose the reference direction that maximizes the received power over all directions. If an optimal direction, $\mu^*$, happens to be one of the reference directions, $\mu_i$, $-P\,{\leq}\,i\,{\leq}\,P$, then there is no optimality gap at all and the URD algorithm achieves the optimal solution. Moreover, as $D$ increases, the gap between an optimal direction and its closest reference direction decreases, and intuitively speaking, as $D\,{\rightarrow \infty}\,$, one of the reference directions coincides with an optimal direction.

\section{Results and Discussions} 

\subsection{Simulation Setup} 
Extensive simulations were conducted to evaluate the performance of the proposed URD algorithm under different system configurations. For each configuration, 1000 distinct channel realizations were considered. The number of reflecting elements, $N$, was varied from $64$ to $1024$, and the number of reference directions, $D$, was varied over $\{1,3,5,7,9,11,13\}$. The simulations were conducted for $K=2$ and $K=4$ discrete phase shifts. 

The proposed URD algorithm was compared with the Closest-Point Projection (CPP)~\cite{zhou2023}, the Approximation (APX) algorithm~\cite{zhang2022} and Algorithm 3 of~\cite{pekan2024a}. The proposed URD algorithm with $D\,{=}\,1$ reduces to the ARB algorithm of \cite{pavithta2025} by algorithm descriptions. We numerically show that the optimality gap of the proposed URD algorithm with $D\,{=}\,1$ is identical to that of the CPP algorithm of \cite{zhou2023}. When $D\,{=}\,3$, the three uniformly spaced reference directions coincide with the three sector centers used in APX; consequently, the resulting optimality gaps are identical. 

%The proposed URD algorithm was compared with the following baseline methods:
%
%\begin{itemize}
%    \item {Closest-Point Projection (CPP)~\cite{zhou2023}:}
%    A sub-optimal method that selects discrete phase shifts based on
%    the closest-point projection criterion.
%
%    \item {Approximation (APX) algorithm~\cite{ren2023}:}
%    A sub-optimal method that uses a three-sector approximation to
%    determine the discrete phase shifts. 
%
%    \item {Algorithm 3 of~\cite{pekan2024a}:}
%    An optimal algorithm proposed for discrete-phase RIS optimization.
%    This algorithm serves as the benchmark for evaluating the
%    optimality gap and computational efficiency of URD and the other
%    sub-optimal methods. It achieves the optimal received power using
%    elementwise updates and considers the candidate configurations
%    required to identify the optimum.
%\end{itemize}

%The Closest-Point Projection (CPP) and three-sector Approximation (APX) methods are included as reference methods to establish the relationship between URD and conventional approaches. 

The higher-direction configurations ($D\,{=}\,{5,7,9,11,13}$) are used to investigate the effect of increasing the number of directions on optimality and computational
efficiency. The worst-case optimality gap and speed-up ratio are used as the performance metrics, with Algorithm~3 of~\cite{pekan2024a} serving as the optimal benchmark. 

\subsection{Performance Results} 
Fig.~3(a) compares the worst-case optimality gaps of the URD algorithm with those of CPP and APX for $K\,{=}\,4$. The equivalence of the optimality-gap curves obtained by CPP, APX, and URD with $D\,{=}\,1$ and $D\,{=}\,3$ demonstrates that these existing approximation methods can be interpreted as particular operating points within the proposed URD framework. Increasing $D$ beyond these operating points systematically reduces the optimality gap, thereby enabling a continuous complexity-performance tradeoff unavailable in existing methods. As $D$ increases beyond these configurations, the worst-case optimality gap generally decreases, indicating that considering more directions enables URD to obtain solutions closer to the optimal benchmark. 

It can be observed that:

\begin{itemize}
    \item[\textbf{O1:}] As shown in Fig.~3(a), Fig.~4(a), and Fig.~5(a),
    the worst-case optimality gap decreases significantly with increasing
    $D$. For $K\,{=}\,2$ and $N\,{=}\,64$ Fig.~4(a), the gap decreases from
    approximately $60\%$ at $D\,{=}\,1$ to below $1\%$ at $D\,{=}\,13$. For $K\,{=}\,4$
    Fig.~5(a), it decreases from approximately $16\%$ to below $1\%$
    over the same range. These results demonstrate that increasing the
    number of directions enables URD to achieve solutions closer to the
    optimal benchmark.

    \item[\textbf{O2:}] Fig.~3(b), Fig.~4(b), and Fig.~5(b) show that
    the speed-up ratio generally increases with $N$ for a fixed $D$.
    For $K\,{=}\,4$, the speed-up of the $D\,{=}\,1$ configuration increases from
    approximately $20\times$ at $N=64$ to $28\times$ at $N\,{=}\,1024$
    Fig.~3(b). This shows the increasing computational advantage of the proposed URD algorithm as the size of RIS increases.

  \item[\textbf{O3:}] The optimality plot shows that the optimality gap approaches zero as $D$ increases. For $K\,{=}\,4$ and $N\,{=}\,1024$, the gap approaches $0.1\%$ at $D\,{=}\,13$, while remaining below $1\%$ across the tested values of $N$. These results demonstrate that the proposed URD algorithm can closely approach the optimal benchmark for sufficiently large $D$.
\end{itemize}

\begin{figure}[tb]
  \centering
  \begin{subfigure}{0.45\textwidth}
    \centering
    \includegraphics[width=\linewidth]{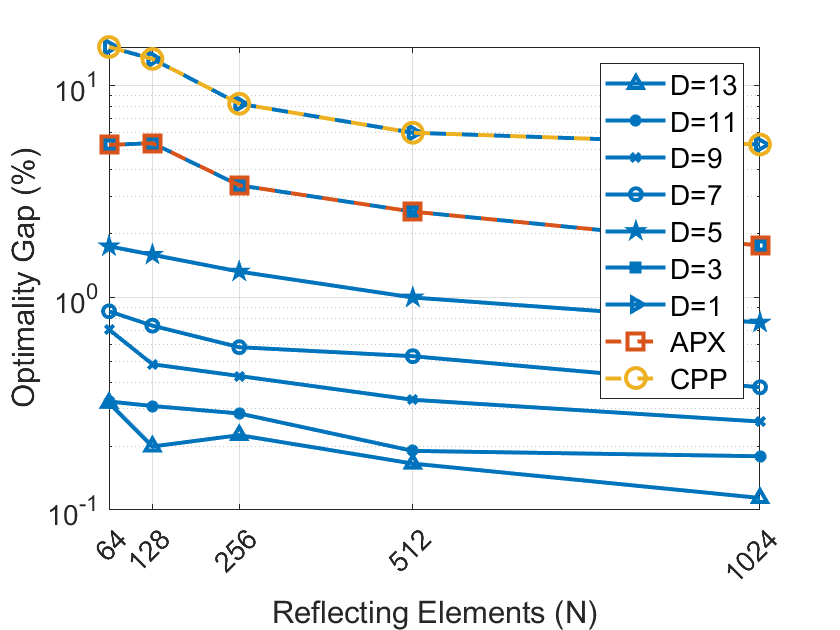}
    \caption{Comparison of Optimality Gap.}
    \label{optigap_N_8}
  \end{subfigure}
  \hfill
  \begin{subfigure}{0.45\textwidth}
    \centering
    \includegraphics[width=\linewidth]{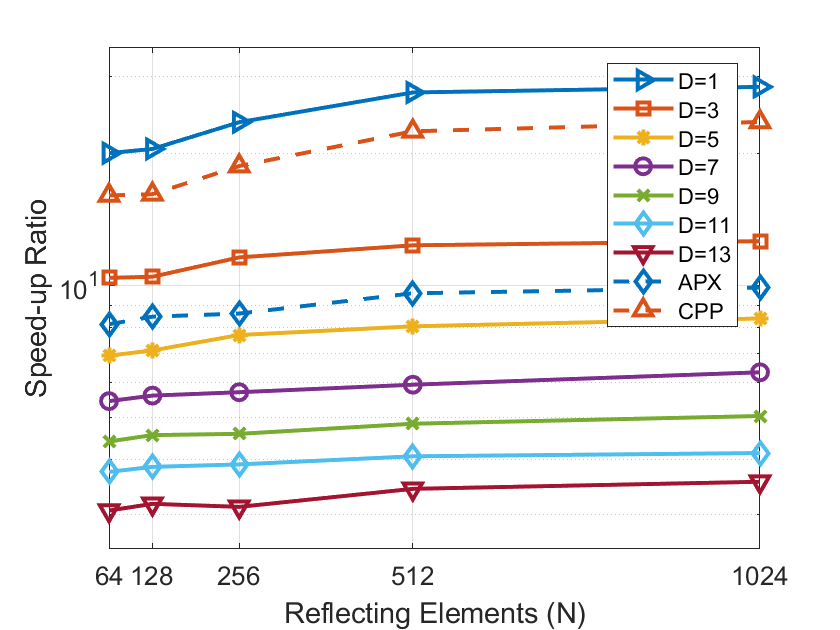}  
   \caption{Comparison of Speedup.}
    \label{speedup_N_8}
  \end{subfigure}
  \caption{Comparison of CPP, APX, and Algorithm 1 with different $D$ values (1,3,5, 7 and 9), based on 1000 channel realizations for an increasing number of REs, $N$, and a fixed number of phase-shifts $K\,{=}\,4$.} 
  \label{N_8}
\end{figure}

\begin{figure}[tb]
  \centering
  \begin{subfigure}{0.45\textwidth}
    \centering
    \includegraphics[width=\linewidth]{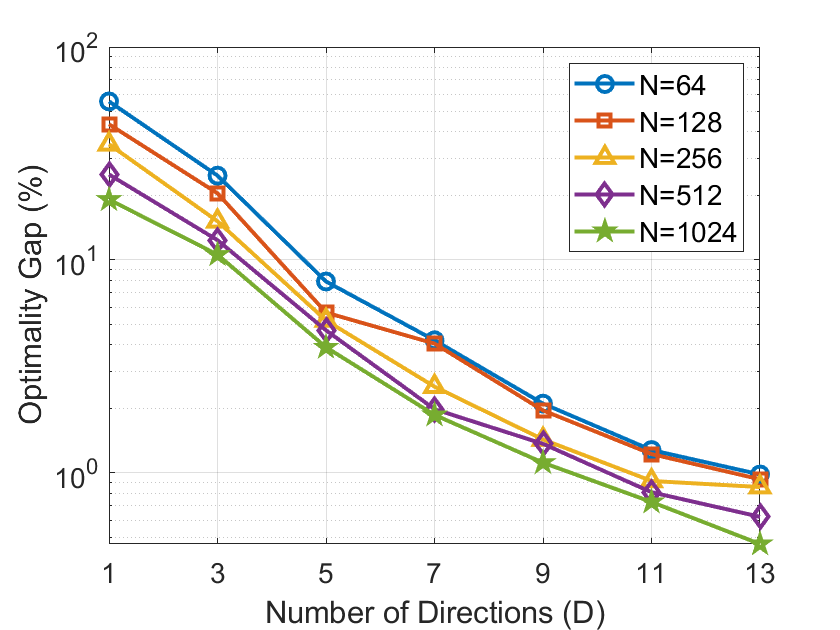}
    \caption{Optimization Gap with $K\,{=}\,2$.}
    \label{opti_varyN}
  \end{subfigure}
  \hfill
  \begin{subfigure}{0.45\textwidth}
    \centering
    \includegraphics[width=\linewidth]{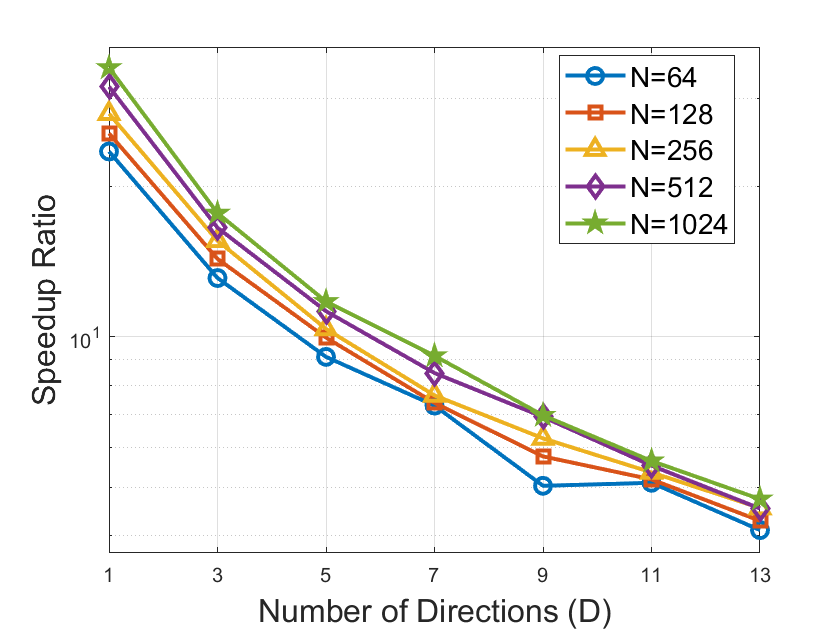}
   \caption{Speedup with $K\,{=}\,2$.}
    \label{speed_varyN}
  \end{subfigure}
  \caption{Comparison of Optimality Gap and Speedup due to the URD algorithm for different number of reference directions, $D$, based on 1000 channel realizations, with $K\,{=}\,2$ and varying reflecting elements $N$.}
  \label{varyN}
\end{figure}

\begin{figure}[tb]
  \centering
  \begin{subfigure}{0.45\textwidth}
    \centering
    \includegraphics[width=\linewidth]{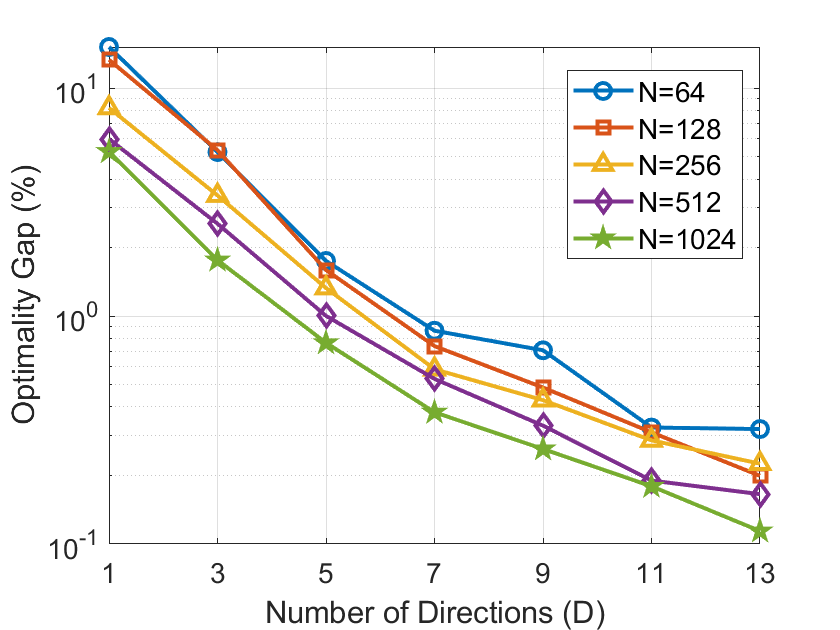} 
    \caption{Optimization Gap with $K\,{=}\,4$.}
    \label{opti_varyN}
  \end{subfigure}
  \hfill
  \begin{subfigure}{0.45\textwidth}
    \centering
    \includegraphics[width=\linewidth]{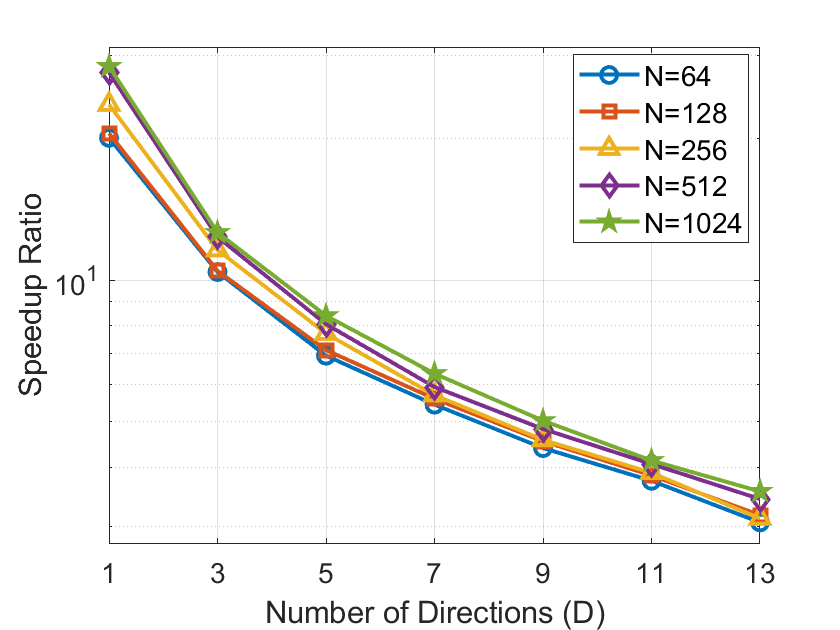}  
   \caption{Speedup with $K\,{=}\,4$.}
    \label{speed_varyN}
  \end{subfigure}
  \caption{Comparison of Optimality Gap and Speedup due to the URD algorithm for different number of reference directions, $D$, based on 1000 channel realizations, with $K\,{=}\,4$ and varying reflecting elements $N$.}
  \label{varyN}
\end{figure}

\section{Conclusion}

This paper presented a low-complexity discrete phase-shift optimization algorithm for RIS-assisted communications that employs multiple reference directions to construct candidate phase configurations. Unlike existing approximation methods that operate at fixed complexity-performance points, the proposed approach introduces a tunable parameter that enables flexible control over the trade-off between computational cost and received-power performance. Numerical results demonstrated that the proposed algorithm achieves substantial computational speedups while maintaining a small optimality gap relative to optimal solutions. Moreover, the optimality gap was shown to decrease systematically as the number of reference directions increases, providing a practical means of tailoring algorithmic performance to application-specific requirements. The proposed algorithm therefore offers an attractive balance between computational efficiency and beamforming performance for large-scale RIS deployments.

%\appendix

%\end{document}

\clearpage
\newpage 
\onecolumn
\appendix

Assuming $N$ to be large, the direct LoS path can be practically discarded. Then, the received power is given by

\begin{equation}
\label{eq:RxPowerApprox}
f\!\left(\boldsymbol{\theta}^{\mathrm{URD}}\right)
\approx
\left|
\sum_{n=1}^{N}
\beta_n e^{j\left(\alpha_n + \theta_n^{\mathrm{URD}}\right)}
\right|^2,
\end{equation}
where $\theta_n^{\mathrm{URD}}$ denotes the optimal $\theta_n$ obtained as per the URD algorithm for the $n^{\text{th}}$ RE. Ignoring the LoS path, we can replace the approximation in \eqref{eq:RxPowerApprox} with an equality and rewrite the received power as 

\begin{align}
f\!\left(\boldsymbol{\theta}^{\mathrm{URD}}\right) &=
\sum_{n=1}^{N} \beta_n^2
+
2 \sum_{k} \sum_{\ell<k}
\beta_k \beta_\ell
\cos\!\left[
(\alpha_k + \theta_k^{\mathrm{URD}})
-
(\alpha_\ell + \theta_\ell^{\mathrm{URD}})
\right]. \\
%\end{align}
%\begin{align}
&=\sum_{n=1}^{N} \beta_n^2
+
2 \sum_{k} \sum_{\ell<k}
\beta_k \beta_\ell
\cos\!\Big[
(\alpha_k + \theta_k^{\mathrm{URD}} - \angle \mu^{\mathrm{URD}})
-
(\alpha_\ell + \theta_\ell^{\mathrm{URD}} - \angle \mu^{\mathrm{URD}})
\Big].
\end{align}
where $\angle \mu^{\mathrm{URD}}$ is the reference direction for which maximum received power is achieved among all tried directions $\angle \mu_{-P} \;\le\; \angle \mu_i \;\le\; \angle \mu_{P}$. Sine the URD algorithm takes the maximum among all the tried directions, we have 

\begin{align}
f\!\left(\boldsymbol{\theta}^{\mathrm{URD}}\right) 
&=\sum_{n=1}^{N} \beta_n^2
+
\max_{\,-P \le i \le P}
\; 2 \sum_{k} \sum_{\ell<k}
\beta_k \beta_\ell
\cos\!\Big[
(\alpha_k + \theta_k^{(i)} - \angle \mu_i)
-
(\alpha_\ell + \theta_\ell^{(i)} - \angle \mu_i)
\Big] \\
\label{eq:Objfn}
&= \sum_{n=1}^{N} \beta_n^2
+
\max_{\,-P \le i \le P}
\; 2 \sum_{k} \sum_{\ell<k}
\beta_k \beta_\ell
\cos(\Delta_k^{(i)} - \Delta_{\ell}^{(i)}), 
\end{align}
where $\Delta_k^{(i)} := \alpha_k + \theta_k^{(i)} - \angle \mu_i$, and $\theta_k^{(i)}$ is chosen such that $\alpha_k + \theta_k^{(i)}$ is aligned as closely as possible to $\angle \mu_i$, the $i^{\text{th}}$ reference direction. 

Note that the right hand side in \eqref{eq:Objfn} depends on $P$, and so let us denote it as $x_P$, which represents the optimal received power as per the URD algorithm obtained by taking a maximum over $2P+1$ evaluations corresponding to $2P+1$ reference directions. Consider the sequence $\{x_P\}$, with $P = 2^j, j = 1, 2, \dots$, where the reference directions pertaining to $x_{2^j}$ are obtained by keeping the reference directions pertaining to $x_{2^{j-1}}$ and adding new reference directions by taking the bisectors of  the reference directions pertaining to $x_{2^{j-1}}$. Then, the sequence $\{x_{2^j}\}$, $j = 1, 2, \dots$, is a monotonically non-decreasing sequence as the maximum is taken over an increasing sequence of sets (in the sense of set-theoretic inclusion) of reference directions. Clearly, therefore, by Monotone Convergence Theorem (MCT), the sequence $\{x_{2^j}\}$, $j = 1, 2, \dots$, converges to the supremum, which, in this case, is the true optimum received power achievable for the given channel gains.

%Define $\Delta_k^{(i)} = \alpha_k + \theta_k^{(i)} - \angle \mu_i$. For a Rayleigh fading channel, and assuming the REs to be spatially separated by distances $d \geq \lambda/2$, where $\lambda$ is the wavelength of operation, we could assume, $\forall i$, that (i) the $\Delta_k^{(i)}$'s are independent for different $k$'s, $\forall i$, (ii) the $\beta_k$'s are independent for different $k$'s, $\forall i$, (iii) the $\Delta_k^{(i)}$'s and $\beta_k$'s are independent for the same $k$, $\forall i$, (iv) the $\Delta_k^{(i)}$'s are uniformly distributed as $\sim \mathcal{U}\!\left[-\frac{\pi}{K},\, \frac{\pi}{K}\right]$, $\forall i,k$, and (v) the $\beta_k$'s are Rayleigh distributed.
%
%\begin{align}
%&=
%\sum_{n=1}^{N} \beta_n^2
%+
%\max_{\,-P \le i \le P}
%\; 2 \sum_{k} \sum_{\ell<k}
%\beta_k \beta_\ell
%\cos\!\left(
%\Delta_k^{(i)} - \Delta_\ell^{(i)}
%\right).
%\end{align}
%
%
%\begin{align}
%E\left[f\!\left(\boldsymbol{\theta}^{\mathrm{URD}}\right)\right] & \geq 
%\sum_{n=1}^{N} \beta_n^2
%+
%\max_{\,-P \le i \le P}
%\; 2 \sum_{k} \sum_{\ell<k}
%\beta_k \beta_\ell
%\cos\!\left(
%\Delta_k^{(i)} - \Delta_\ell^{(i)}
%\right).
%\end{align}

\end{document}